# Scalable production of graphene inks *via* wet-jet milling exfoliation for screen-printed micro-supercapacitors

*Sebastiano Bellani, Elisa Petroni, Antonio Esau Del Rio Castillo, Nicola Curreli, Beatriz Martín-García, Reinier Oropesa-Nuñez, Mirko Prato and Francesco Bonaccorso\**

Dr. S. Bellani, E. Petroni, Dr. A. E. Del Rio Castillo, N. Curreli, Dr. B. Martín-García, Dr. F. Bonaccorso
Graphene Labs, Istituto Italiano di Tecnologia, via Morego 30, 16163 Genoa, Italy
E-mail: francesco.bonaccorso@iit.it
Dr. F. Bonaccorso, Dr. R. Oropesa-Nuñez
BeDimensional Spa., Via Albisola 121, 16163 Genova, Italy
Dr. M. Prato
Materials Characterization Facility, Istituto Italiano di Tecnologia, via Morego 30, 16163 Genoa, Italy

**Abstract**

The miniaturization of energy storage units is pivotal for the development of next-generation portable electronic devices. Micro-supercapacitors (MSCs) hold a great potential to work as on-chip micro-power sources and energy storage units complementing batteries and energy harvester systems. The scalable production of supercapacitor materials with cost-effective and high-throughput processing methods is crucial for the widespread application of MSCs. Here, we report wet-jet milling exfoliation of graphite to scale-up the production of graphene as supercapacitor material. The formulation of aqueous/alcohol-based graphene inks allows metal-free, flexible MSCs to be screen-printed. These MSCs exhibit areal capacitance ($C_{areal}$) values up to 1.324 mF cm$^{-2}$ (5.296 mF cm$^{-2}$ for a single electrode), corresponding to an outstanding volumetric capacitance ($C_{vol}$) of 0.490 F cm$^{-3}$ (1.961 F cm$^{-3}$ for a single electrode). The screen-printed MSCs can operate up to power density above 20 mW cm$^{-2}$ at energy density of 0.064 μWh cm$^{-2}$. The devices exhibit excellent cycling stability over charge-discharge cycling (10000 cycles), bending cycling (100 cycles at bending radius of 1 cm) and folding (up to angles of 180°). Moreover, ethylene vinyl acetate-encapsulated MSCs retain their electrochemical properties after a home-laundry cycle, providing waterproof and washable properties for prospective application in wearable electronics.

# 1. Introduction

Supercapacitors, also known as ultracapacitors or double-layer capacitors, are potential energy storage devices which store charges at the electrode/electrolyte interfaces through rapid and reversible adsorption/desorption of ions.[1,2] Since they do not involve chemical reactions, supercapacitors can be charged quickly (from millisecond- to second-scale), leading to a very high power density (> 10000 W kg$^{-1}$)[3,4,5,6,7] and long cycle life (millions of charge-discharge (CD) cycles).[8,9] In addition, they can be flexible and made entirely of non-toxic and non-hazardous materials,[10,11] fulfilling peculiar requirements of portable and wearable electronic devices.[12,13,14] In this context, the ever-increasing demand of multi-functional systems has drawn the attention towards planar micro-supercapacitors (MSCs),[15,16,17] which can deliver all the benefits of supercapacitors in smaller and lighter devices.[18,19] Furthermore, their interdigitated planar geometry, in contrast to the conventional sandwich structure, provides power densities that are several orders of magnitude larger than those of batteries and vertical supercapacitors,[15,17] offering easy on-chip integration into miniaturized device requiring nano-/micro-scale power supply.[20,21] Currently, the development of nanostructured electroactive materials[22,23,24,25,26] and thin-film manufacture technologies[15,27,28] is pivotal to produce MSCs, which provide both ultrahigh power density and high-energy density (approaching those of thin-film batteries, between $10^{-3}$–$10^{-2}$ Wh cm$^{-3}$).[29,30] However, scaling-up the production of electroactive materials is still challenging.[31] Moreover, the fabrication of MSCs with active material mass loadings and capacities comparable to those of the vertical counterparts[15-19] is challenging too. Additionally, it is necessary to adopt industrial scalable techniques for MSC device fabrication.[32,33] With regards to the active materials, carbon is a viable and inexpensive material, and its nanostructured allotropes, such as activated carbon,[34,35] carbon nanotubes,[36,37,38] carbide-derived carbons[39] and carbon nano-onions,[40] have been reported as active materials for MSCs. Nevertheless, the limited ion accessibility into the activated carbon[41,42] sets mass loading and power restrictions,[35,43,44] while the non-scalable production of other nanocarbons,[15-19] as well as the complexity of their film patterning,[15-19] are key technological shortcomings. In this context, graphene is a promising active material for MSCs,[45,46] (and for all supercapacitors in general[47,48,49,50,51,52,53,54]) because of its high theoretical specific surface area -SSA- (2630 m$^2$ g$^{-1}$)[45-55,56] and excellent electrical conductivity (*e.g.*, charge carrier mobility of ~5000 cm$^2$ V$^{-1}$ s$^{-1}$).[45-56] Its theoretical specific capacitance (550 F g$^{-1}$)[57] enables graphene-based supercapacitors to achieve superior energy densities compared with those of other carbon-based supercapacitors,[58,59,60] including activated carbons (ACs) (in the order of 100 F g$^{-1}$ due to the lack of efficient ion transport in presence of "closed micropores"[61,62,63]), carbon nanotubes (CNTs) (< 200 F g$^{-1}$ due to limited SSA,[64,65,66] *e.g.*, 1315 m$^2$ g$^{-1}$ for single-walled carbon nanotubes –SWCNTs,[67,68] ~800 m$^2$ g$^{-1}$ for double wall CNTs,[67,69] ~50 m$^2$ g$^{-1}$ for 40-wall CNTs[67,69]) and carbide-derived carbons (~200 F g$^{-1}$ using ionic liquid electrolytes[70,71], typically < 180 F g$^{-1}$ using organic electrolytes[72,73,74]). Lastly, in MSCs, the in-plane displacement of graphene is parallel to the ions movements during the device charging/discharging. Consequently, MSC configuration intrinsically facilitates the flow of the electrolyte ions between the graphene layers in a short diffusion pathway,[32,75,76,77,78] maximizing the ion accessibility to the active material surface area.[32,75,78]

Among the method for the production of graphene, liquid phase exfoliation (LPE)[79,80,81,82] is a top-down approach, which allows the bulk pristine graphite to be exfoliated into graphene in a liquid environment. This process is performed by exploiting ultrasound[83,84,85,86,87,88,89,90] or shear forces[91,92,93,94,95] to break the van der Waals bond between the adjacent planes of the graphite.[79-82] The possibility to produce graphene in a liquid phase allows functional inks with on-demand rheological and morphological properties to be formulated.[80-82,96,97,] This represents a step forward for the development of industrial-scale, reliable and inexpensive printing/coating processes, which represents a boost for the full exploitation of such nanomaterial.[81,82,98] Among the LPE techniques, ultrasonication is the most investigated one because of its easiness.[79-82] Actually, ultrasonication-induced exfoliation is largely due to cavitation.[99] In fact, collapsing cavitation bubbles, tensile and shear stresses provide the energy input to overcome the graphite intersheet interactions, leading to both exfoliation and fragmentation.[79-82,98,100] Unfortunately, the scalability of ultrasonication method has not been demonstrated yet. The time required to obtain 1 g of exfoliated material ($t_{1gram}$), the volume of solvent required to produce 1 g of exfoliated material, ($V_{1gram}$), as well as the ratio between the weight of the final graphitic material and the weight of the starting graphite flakes (defined as exfoliation yield –Y–), are still insufficient for industrial-scale productions.[101,102,103] For example, after more than 35 h of sonication and several steps of sonication and re-dispersion, the concentration of the processed material can reach 60 g L$^{-1}$, which means a $t_{1gram}$ > 1800 min, a $V_{1gram}$ = 0.53 L and a Y of 19%.[101] For a 6 h of sonication, a $t_{1gram}$ > 180 min, a $V_{1gram}$ = 3.3 L and a Y = 3% were attained.[101] As alternative LPE technique, mechanical exfoliation of graphite *via* ball-milling also requires long processing times,[83,104,105] resulting in a $t_{1gram}$ = 60 min, a $V_{1gram}$ = 100 L and a low Y (< 1%).[83] Recently, the use of shear mixer,[106,107] micro-fluidization[108] or jet stream mill[109] have also been reported for the exfoliation of industrial volume of graphite. However, shear mixing provides high $t_{1gram}$ (~3600 min), and extremely low Y (~0.002%),[110,111] while micro-fluidization/jet stream mill produce highly defective exfoliated materials mainly composed by thick flakes (4-70 nm).[108],[109] Therefore, these exfoliation techniques are still unattractive for scaling-up practical application, including MSCs.

Recently, our laboratories developed a novel approach for the exfoliation of graphite (as well as other layered crystals), based on high-pressure wet-jet-milling (WJM),[112,113] resulting in a production of 2 L h$^{-1}$ of defect-free and high quality single- and few-layer graphene flakes highly concentrated (10 g L$^{-1}$) dispersion in N-methyl-2-pyrrolidone (NMP), making the scaling-up more affordable ($t_{1gram}$ = 2.55 min, $V_{1gram}$ = 0.1 L, Y ~100%).[112] Based on these results, in this work we exploited WJM process to produce single-/few-layer flakes of graphene (WJM-graphene) as active material in supercapacitor electrodes (**Figure 1**a). The main advantage of WJM compared to all the LPE techniques, is the process time of the sample, *i.e.*, the passage of the processed dispersion through the nozzle, which is reduced to a fraction of a second, instead of hours in a sonic bath or shear exfoliation.[112] A simple solvent-exchange process[114,115,116] was carried out to re-disperse the WJM-graphene in water/ethanol ($H_2O$/EtOH) (70:30) and terpineol (1% wt.), providing a screen printable ink to fabricate solid-state MSCs on flexible substrate (polyethylene terephthalate –PET–) (Figure 1b). The addition of SWCNTs dispersed in EtOH to the WJM-graphene ink (Figure 1c) and the use of pyrolytic graphite (PG) paper as current collector (Figure 1d) were used for increasing the active film porosity and decrease the current collector resistance, respectively. Noteworthy, CNTs have been previously adopted as spacers between the graphene flakes to avoid restacking effects of graphene flakes during their deposition[51,117] and increasing both surface area and pore size of supercapacitor electrodes.[118,119] The printed MSCs exhibited areal capacitance ($C_{areal}$) values up to 1.324 mF $cm^{-2}$ (5.296 mF $cm^{-2}$ for a single electrode), corresponding to an outstanding volumetric capacitance ($C_{vol}$) of 0.490 F $cm^{-3}$ (1.961 F $cm^{-3}$ for a single electrode). The screen-printed MSCs have shown high-rate capability (power density above 20 mW $cm^{-2}$ at energy density of 0.064 µWh $cm^{-2}$), as well as excellent cycling stability over charge-discharge cycling (10000 cycles), bending cycling (100 cycles at bending radius of 1 cm) and folding (up to angles of 180°). Moreover, ethylene vinyl acetate (EVA)-encapsulated MSCs retained their electrochemical properties after a home-laundry cycle, providing waterproof and washable properties for prospective application in wearable electronics.

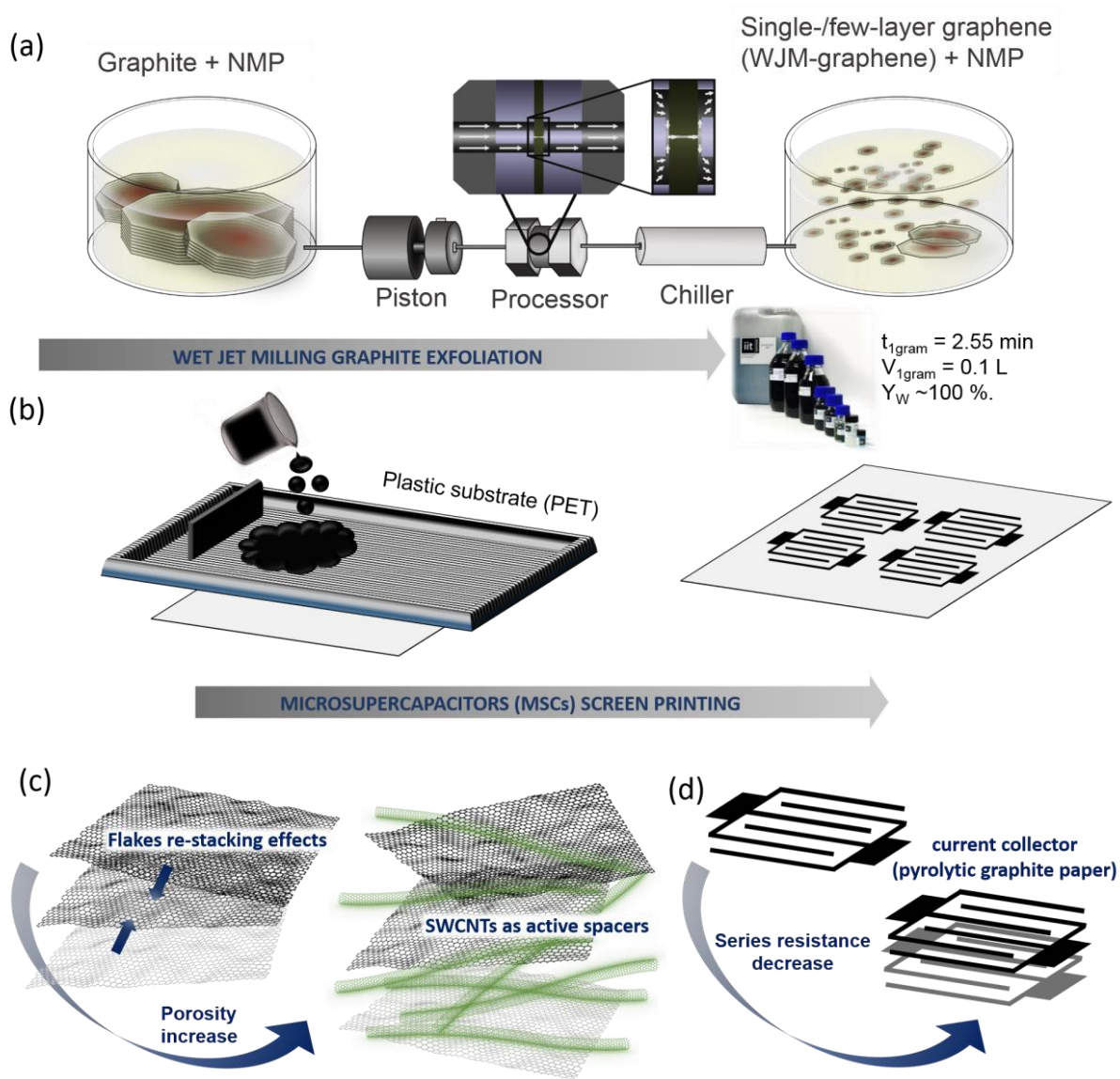

**Figure 1.** (a) Schematic illustration of the production of single-/few-layer graphene by WJM exfoliation of graphite (WJM-graphene). (b) Screen printing of MSCs onto plastic substrate (PET). (c) Addition of SWCNTs as active spacers for avoiding the re-stacking of the flakes. (d) Use of pyrolytic graphite (PG) paper in order to decrease the current collector resistance of MSCs for high-power density requirements.

## 2. Results and discussion

## 2.1. Wet jet milling graphene production and characterization

The WJM apparatus, as schematically illustrated in Figure 1a, makes use of a high-pressurized jet stream to homogenize and exfoliate the sample, *i.e.*, a layered material.[112,113,120] More in detail, a hydraulic mechanism and a piston supply the pressure in order to direct the mixture of solvent and layered crystals into the processor, where the generation of shear forces promotes the sample exfoliation.[112,113,120,121] Additional details of the WJM process, including the description of the processor, is reported in Supporting Information (**Figure S1**). The time during which the layered materials are subjected to exfoliation is less than one second.[112,120] Immediately after the processor, the sample is cooled down by means of a chiller. The WJM process can be repeated in cascade by adjusting the dimension of the nozzle (from 0.3 mm to 0.1 mm) in order to optimize the exfoliation process and to finely tune the morphology of the resulting flakes. In this work, we processed graphite through three WJM passes (see additional details in Experimental section), and the as-prepared WJM-graphene was then investigated as active material for MSCs. The optimization of the WJM process through multiple passes has been previously discussed in a recent work,[112] in which it has been demonstrated a $t_{1gram}$ = 2.55 min, a $V_{1gram}$ = 0.1 L and a Y of ~100 %. These are important values because they satisfy the request of active material (in terms of amount of material and volume) for the development of MSCs on industrial scale.[112]

The lateral size and thickness of the as-produced WJM-graphene were characterized by means of transmission electron microscopy (TEM) (**Figure 2**a,b) and atomic force microscopy (AFM) (Figure 2c,d). The sample consisted of irregularly shaped (Figure 2a) and nm-thick flakes (Figure 2c). Statistical analysis indicated that the lateral size and thickness of the flakes followed lognormal distributions, peaked at ~460 nm (Figure 2b) and ~3.2 nm (Figure 2c), respectively.

Raman spectroscopy characterization was carried out in order to evaluate the structural properties and the quality of the as-produced WJM-graphene. A typical Raman spectrum of graphene shows, as fingerprints, G, D and 2D peaks (see Supporting Information for more details).[122,123] For single-layer graphene the 2D band is roughly four times more intense than the G peak.[122] Multi-layers graphene (> 5 layers) exhibits a 2D peak, which is almost identical, in term of both intensity and lineshape, to the graphite case (intensity of the $2D_2$ band is twice the 2D$_1$ band).[124,125] Few-layers graphene (< 5 layers), instead, has a 2D$_1$ peak more intense than the 2D$_2$.[126] Taking into account the intensity ratios of the 2D$_1$ and 2D$_2$, it is possible to roughly estimate the flake thickness.[51,112, 127, 128, 126] Figure 2e shows the comparison between the Raman spectra (normalized to the G peak intensity) of the graphite and WJM-graphene. The increase in intensity of the D and D' peaks in the Raman spectrum of the WJM-graphene is related to a decrease on the flake size.[129,130,131,132] In particular, Raman statistical analysis shows that the ratio between the intensity of D and G peak (I(D)/I(G)) ranges from 0.1 to 1.2 for WJM-graphene (**Figure S2**a), whereas approaches to null values in graphite. For WJM-graphene, the position of G peak (Pos(G)) and the full width half maximum of the G band (FWHM(G)) range from 1578 to 1583 cm$^{-1}$ (Figure S2b) and from 14 to 25 cm$^{-1}$ (Figure S2c), respectively. The plot of I(D)/I(G) *vs*. FWHM(G) (Figure S2d) shows the absence of any correlation, which means that in-plane defects were not originated during the WJM process.[133,134] Figure S2e shows the plot of the intensity of 2D$_1$ – I(2D$_1$)– *vs*. the intensity of 2D$_2$ –I(2D$_2$) (normalized to the I(G)). The dashed line reports the relation I(2D$_1$)/I(G) = I(2D$_2$)/I(G), which represents the multilayer (~5 layers) condition.[122, 135] Thus, graphitic samples whose I(2D$_2$)/I(2D$_1$) value falls above the aforementioned line can be assumed to be less than 5 layers thick, while for the ones that the aforementioned value falls below, are considered thicker than 5 layers, displaying graphite-like properties.[124,136] The results of Figure S2e indicate that WJM-graphene was enriched in single-/few-layer graphene flakes, in agreement with AFM measurements. X-ray photoelectron spectroscopy (XPS) measurements were also carried out on WJM-graphene to ascertain its chemical composition. Figure 2f reports the C 1s spectrum of WJM-graphene, which can be decomposed into different components. The main component peaking at 284.4 eV is referred to sp$^2$ carbon with the corresponding feature due to π-π* interactions at 290.8 eV. A second component, centred at 284.8 eV, refers to the sp$^3$ carbon and is due to flake edges and solvent residuals, as well as to environmental contaminations (adventitious carbon).[137] Two other weak contributions, equal to ~10 % of the total carbon amount, can be ascribed to C-N and C=O groups (peaks at binding energies of 286.3 eV and 287.7 eV, respectively).[138,139] Like sp$^3$ carbon, these nitrogen and oxygen groups come from residual NMP molecules.[112,115,139] These results confirm that high-quality graphene flakes were effectively obtained by WJM.

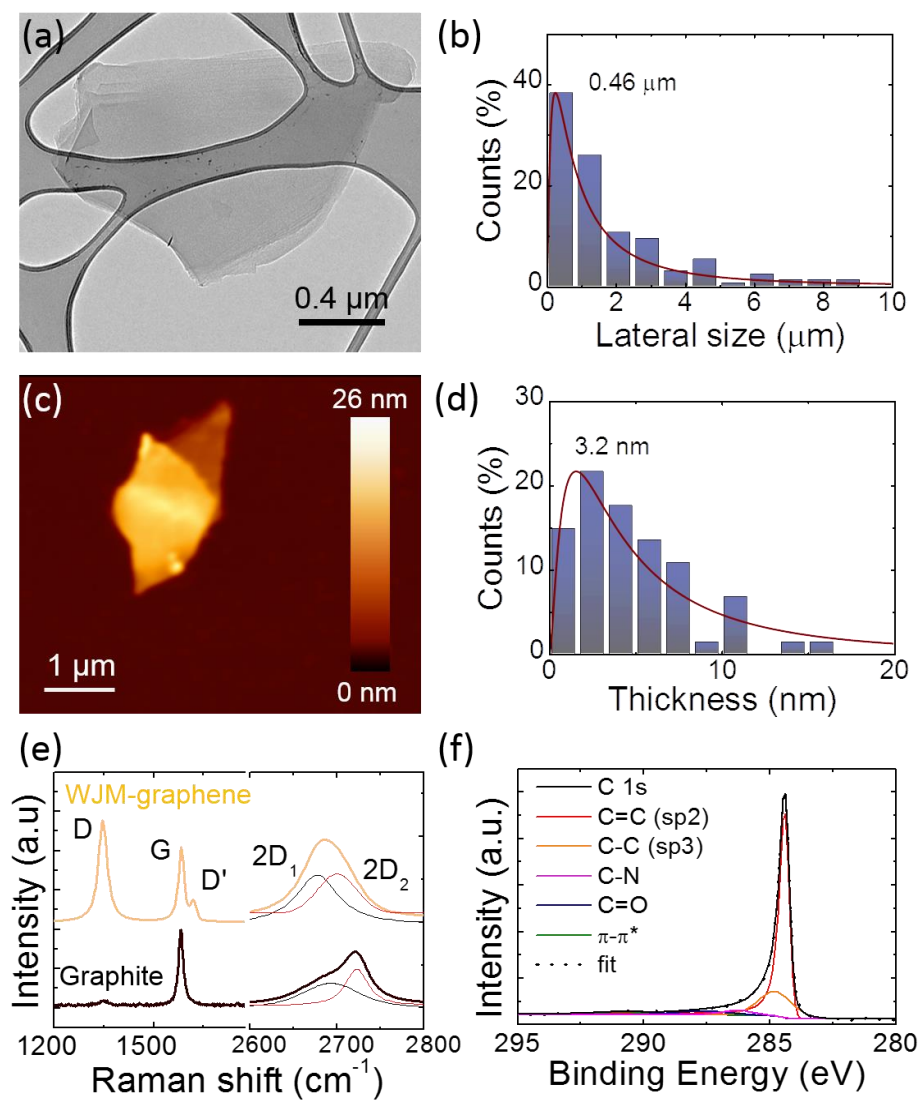

**Figure 2. Morphological, structural and chemical characterization of the WJM-graphene.** (a) TEM image and (b) statistical analysis of the lateral dimension of WJM-graphene (acquired on 170 flakes). (c) Representative AFM image and (d) statistical AFM analysis of the thickness of the WJM-graphene (acquired on 80 flakes). (e) Comparison between the Raman spectra of the graphite (black) and WJM-graphene (orange), with their multi-peak Lorentzian fitting showing the contribution of the individual modes (black line: 2D$_1$; red line: 2D$_2$). (f) C 1s XPS spectrum of the WJM-exfoliated. Its deconvolution is also shown (dashed black line), evidencing the bands ascribed to: C=C (red line), C-C (orange line), C-N (magenta line), C=O (navy line), π-π* (olive line).

The SWCNTs were prepared by dispersing commercial SWCNTs in EtOH at a concentration of 10 g L$^{-1}$ using ultrasonic bath-assisted method for the de-bundling process.[140] Additional details concerning the SWCNTs dispersion production, which follows similar protocols previously reported,[141,142,143] are described in the Experimental Section. As previously shown in our recent work by combing TEM and Raman spectroscopy measurements,[141] the length of the SWCNTs is between 5–30 μm, while the average diameter is less than 1 nm.

**2.2. Micro-supercapacitors screen printing and characterization**

The as-produced WJM-graphene flakes were used as active material for MSCs produced by screen printing on a PET substrate, which was selected as a flexible and cheap substrate. The screen printing ink formulation requires a careful tailoring of the viscosity and surface tension, in order to provide a non-Newtonian fluid with pseudo-plastic and thixotropic properties.[144,145] The latter allow the ink to be optimally transferred onto the substrates, with the ink flowing when sheared by the squeegee, and minimal spreading once printing.[144,145] Furthermore, the ink diluent must be sufficiently volatile in order to facilitate the drying and curing processes of the printed device (allowing optimal process productivity, *i.e.,* high profitability, to be achieved),[144,145] retaining the ink viscosity[146] during the printing and avoiding the so-called "drying-in effect" (*i.e.*, the drying of the ink in the mesh).[147] In order to achieve a screen printable ink, the as-produced WJM-graphene was dried and re-dispersed in a mixture of H$_2$O/EtOH (70:30) and terpineol (weight percent –wt%– of 1) (by solvent-exchange process)[114-116] with a concentration of 75 g L$^{-1}$. Terpineol was used to tune the ink's surface tension and viscosity to suitable values for printing on plastic substrates,[148] in agreement with previous reports on graphene-based inks.[149,150,151,152,153] The addition of SWCNTs into WJM-graphene ink, with a material wt% of 25, was also evaluated in the formulation of the active ink. In fact, it has been demonstrated the CNTs can act as spacers in between the graphene flakes, avoiding restacking effects of graphene flakes during their deposition and increasing the specific area of the resulting films.[51,117] Moreover, CNTs also increases the pore size of graphene-based electrodes.[118,119] Among CNTs, SWCNTs were chosen due to their superior SSA (*i.e.*, theoretical specific capacitance) (~1315 m$^2$ g$^{-1}$[67,68]) in comparison with those of multi-walled CNTs (~800 m$^2$ g$^{-1}$ for double wall CNTs[67,69] and ~50 m$^2$ g$^{-1}$ for 40-wall CNTs[67,69]). Active inks of only SWCNTs were not considered because of the high cost of SWCNTs (~100 USD g$^{-1}$)[154] as well as the low density of the resulting films, which is not desiderable for high-$C_{vol}$ electrodes.

The screen-printed MSCs configuration consisted of 6 fingers (1 mm thick) forming an interdigitated structure (interspace between fingers of 600 μm) with an active area of 1 cm$^2$ (**Figure 3**a). The as-produced MSCs are herein named with the name of the active materials (*i.e.*, WJM-graphene or WJM graphene:SWCNTs). The mass loadings of active materials over the entire MSCs layout area were 1.5 and 1 mg for WJM-graphene and WJM-graphene:SWCNTs, respectively. The mass loading of active material for WJM-graphene was higher than that of WJM-graphene:SWCNTs since a device failure was observed in interdigitated electrodes made of only WJM-graphene with a mass loading of 1 mg. This was a consequence of the excessive electrical resistance (> 100 kΩ, as discussed later in the text) of the electrode. By adding SWCNTs, the electrical resistance can significantly decrease since SWCNTs randomly bridged the graphene flakes to form a more electrically conductive network.[160] Electrodes of PG with the same interdigitated shape of screen-printed electrodes were also evaluated as low-electrical resistance (~0.050 Ω sq$^{-1}$[155]) current collectors for WJM-graphene:SWCNTs-based MSCs. The corresponding MSCs are herein named PG/WJM-graphene:SWCNTs. Figure 3b-e show representative top-view and cross-sectional SEM images of a PG/WJM-graphene:SWCNTs. In particular, Figure 3d-e show that the thickness of the layered structure of the electrode formed by WJM-graphene:SWCNTs onto PG paper is ~27 ± 4 μm. The morphology of the WJM-graphene:SWCNTs film consisted of a mesoporous network of WJM-graphene with dispersed SWCNTs acting as linkers, as well as spacers, between the graphene flakes. **Figure S3** shows a top-view SEM image of a WJM-graphene film in absence of SWCNTs. The SSA of the as-produced films was estimated by Brunauer, Emmett and Teller (BET) analysis of physisorption measurement with Kr at 77 K.[156] The BET SSA of WJM-graphene films was beetween 40–60 $m^2$ $g^{-1}$, while increased up to ~260 $m^2$ $g^{-1}$ for WJM-graphene:SWCNTs film. The low BET SSA of WJM-graphene films can be attributed to the restacking of the single-/few-layer graphene flakes during their film deposition.[51,117] The lower BET SSA of WJM-graphene:SWCNTs films compared to the theoretical ones of graphene and SWCNTs is ascribed to the residual restacking of WJM-graphene flakes, as well as their few-layer morphology.

The MSCs were then completed by coating the printed electrodes with a hydrogel-polymer electrolyte, *i.e.*, poly (vinyl alcohol) (PVA) doped with $H_3PO_4$ (Figure 3a). It is noteworthy that PVA/$H_3PO_4$ mixture has been identified as one of the best solid-state electrolytes for graphene-based supercapacitors.[152, 157] Moreover, compared with traditional liquid electrolytes, hydrogel-polymer electrolytes are preferred due to their higher safety and adaptability to the design of flexible energy storage devices.[158] The possibility to avoid robust metal-based packaging materials also reduces the MSCs thickness for achieving high volumetric performance.[158] This allows the entire fabrication process to be simplified compared to those of conventional vertical supercapacitors.[32,75]

Figure 3f shows a digital photograph of bended MSCs, evidencing the mechanical flexibility of the device.

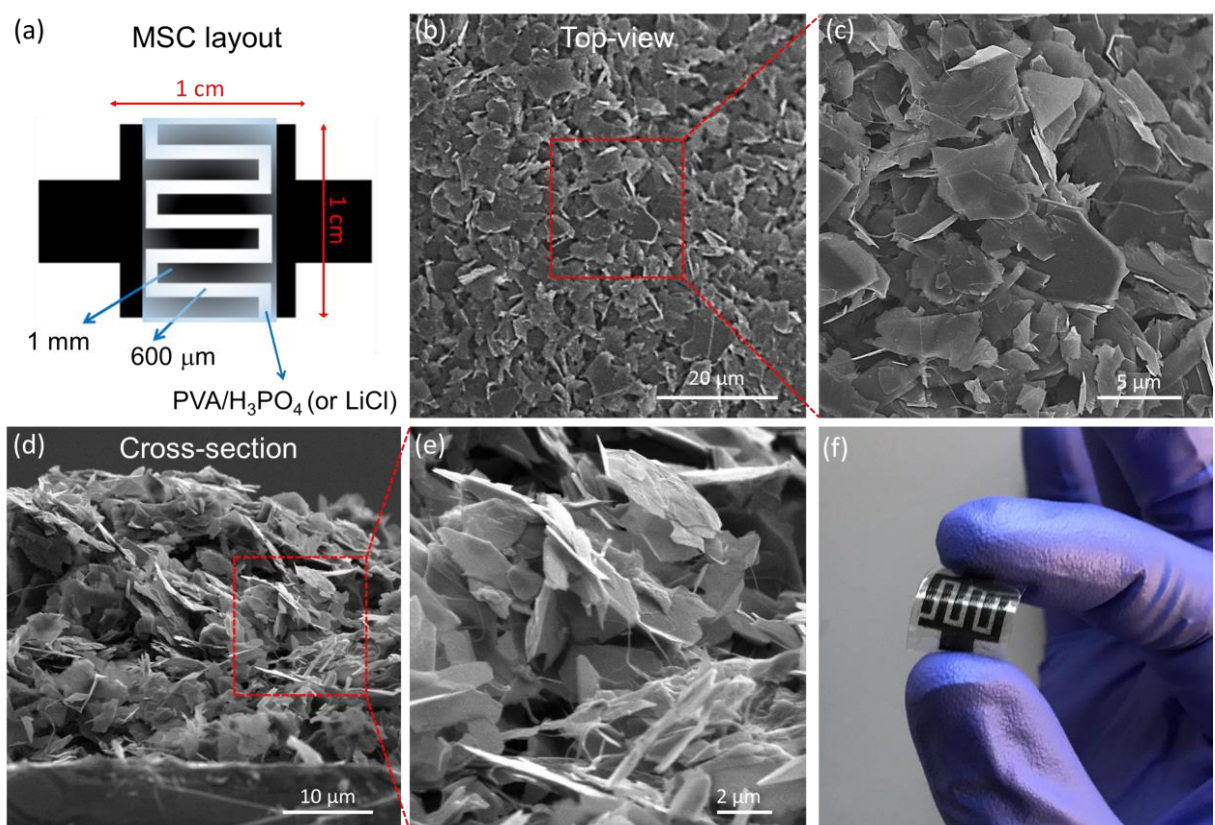

**Figure 3. Structural and morphological characterization of screen-printed MSCs.** (a) Layout adopted as interdigitated electrodes for screen-printing MSCs onto PET substrates. (b) Top-view SEM image, and (c) the corresponding magnification of the region outlined by red dashed lines in b, of a representative PG/WJM-graphene:SWCNTs. (d) Top-view SEM image, and (e) the corresponding magnification of the region delimited by red dashed lines in d, of a representative PG/WJM-graphene:SWCNTs. (f) Digital photograph of a screen-printed MSC. The electrode was manually bended in order to show its mechanical flexibility.

The electrochemical performance of the screen-printed MSCs was evaluated by cyclic voltammetry (CV) (**Figure 4**) and galvanostatic CD measurements (**Figure 5**). Figure 4a shows the comparison between the CV curves of the different MSCs using PVA/$H_3PO_4$ as hydrogel-polymer electrolyte. The nearly-rectangular CV shapes and the absence of redox peaks indicate that the electrodes show double-layer capacitive behaviour[8,9,159] in the investigated voltage ranges. The voltage ranges for each type of MSCs were selected in order to avoid "parasitic" redox reactions (see Supporting Information, **Figure S4**). Clearly, the addition of SWCNTs to the WJM-graphene remarkably increases the area of the voltammograms, which qualitatively indicates an increase of capacitance compared to that of WJM-graphene. In agreement with previous literature,[51,117,119,160,161,162,163] this effect can be attributed to the key-role of SWCNTs that avoids the re-stacking of WJM-graphene flakes during the screen-printing process, which allows the surface area of the flakes to be accessible to the ions for the electrochemical double-layer formation. In addition, the use of PG paper-based current collectors eliminates a biconvex, lens-like shape of the voltammograms, which is instead featured by WJM-graphene and WJM-graphene:SWCNTs. The lens-shaped voltammograms of MSCs without current collectors can attributed to the higher in-plane electrical resistivity of mesoporous WJM-graphene and WJM-graphene:SWCNTs films (~0.8 and ~0.1 $\Omega$ cm, respectively) compared to that of PG paper (~$10^{-5}$ $\Omega$ cm). This data is in agreement with four-point probe measurements shown in Figure 3b. In fact, un-patterned WJM-graphene and WJM-graphene:SWCNTs films with mass loading of 2 mg cm$^{-2}$ (comparable to that used for screen-printed MSCs) displayed sheet resistances as high as 350 and 50 $\Omega$ sq$^{-1}$, respectively, while PG paper has shown a sheet resistance of 0.042 $\Omega$ sq$^{-1}$. The use of PG-based current collectors also enabled the maximum operating voltage to be extended from 1.4 V to 1.8 V. Actually, the efficient charge transport avoids inhomogeneous electrons/ions distributions within nanoporous electrodes,[164, 165] avoiding the formation of localized field which could result in "parasitic" redox-reactions.[165,166] Figures 4c,d report the CV measurements for PG/WJM-graphene:SWCNTs at voltage scan rates ranging from 0.01 to 20 V s$^{-1}$, thus including both low- and high-rate operating conditions. The devices exhibited a voltage scan rate linear dependence of the maximum current density (Figure 4d), indicating that they retained excellent capacitance at voltage scan rate as high as 20 V s$^{-1}$. Notably, such rate performances were achieved for MSCs with μm-thick active material, which are three order of magnitude higher than those reported for high-power density graphene-based MSCs, where the absolute current is significantly inferior compared to our MSCs.[75-78] **Figure S5** reports the CV curves of WJM-graphene and WJM-graphene:SWCNTs, thus in absence of current collector, at different voltage scan rate. As previously discussed, MSCs using solely WJM–graphene were obtained by adopting an active material mass loading which was 50% higher than that of WJM-graphene:SWCNTs. In fact, such mass loading avoided excessive electrical resistance of the WJM–graphene-based interdigitated electrodes, which would otherwise cause the failure of the devices (*i.e.*, resistive-like behaviour).

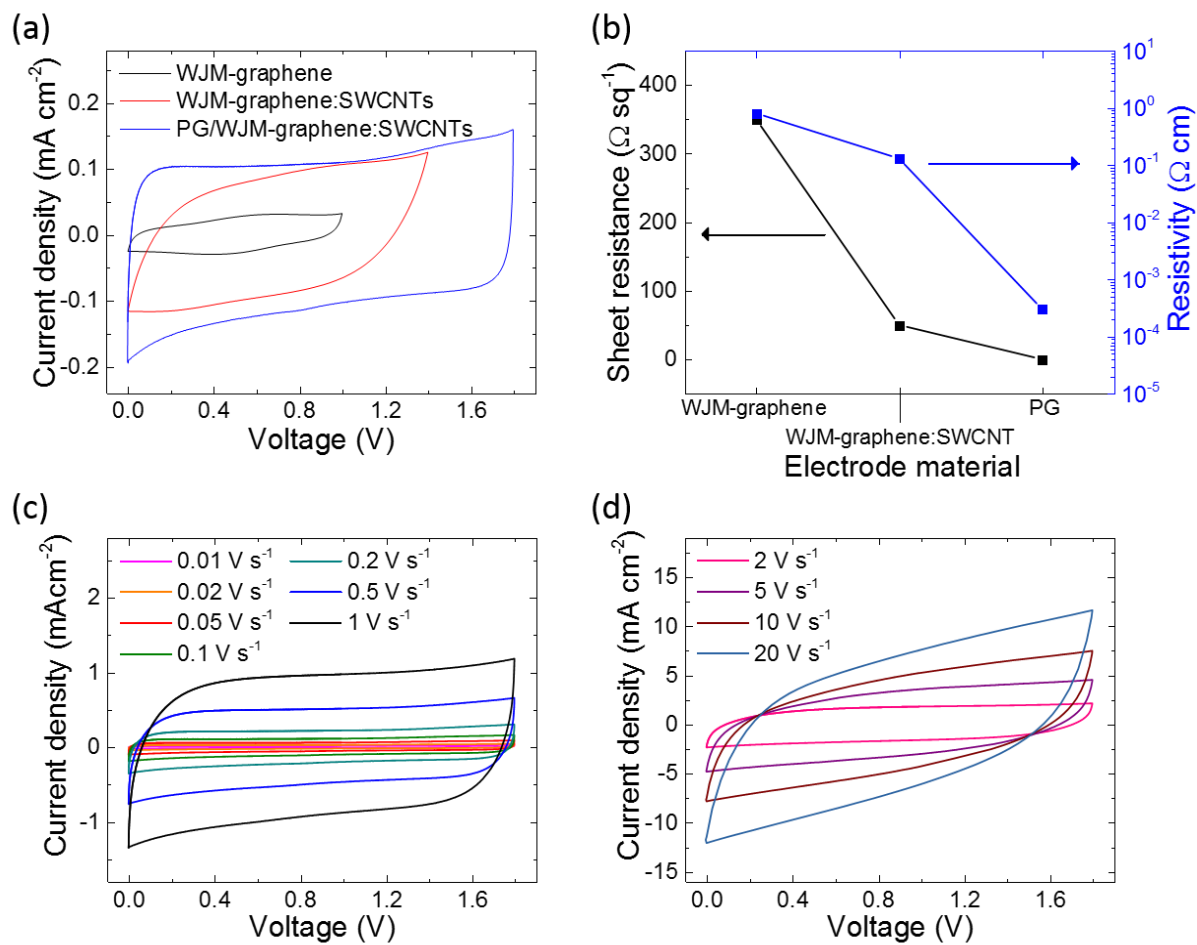

**Figure 4. Electrochemical characterization of screen-printed MSCs.** (a) Comparison between CV curves of WJM-graphene (black), WJM-graphene:SWCNTs (red) and PG/WJM-graphene:SWCNTs (blues) at a voltage scan rate of 100 mV s$^{-1}$. (b) Sheet resistance (left y-axis, black) and resistance (right y-axis, blue) of screen-printed WJM-graphene and WJM-graphene:SWCNT films (thickness of 27 ± 4 and 23 ± 3 μm, respectively, active material mass loading of 2 mg cm$^{-2}$), as well as of PG paper (10 μm-thick). (c, d) Cyclic voltammetry curves of PG/WJM-graphene:SWCNTs at different voltage scan rate (from 0.01 V s$^{-1}$ to 1 V s$^{-1}$ for panel c, from 2 V s$^{-1}$ to 20 V s$^{-1}$ for panel d).

Figure 5a displays the comparison between the CD curves of the different MSCs at current density of 0.1 mA cm$^{-2}$. The MSCs show nearly triangular shapes in the selected voltage ranges. This confirms the double-layer capacitive behaviour of the devices,[8,9,159] which is consistent with the results obtained by CV measurements. Figure 5b shows the CD curves of PG/WJM-graphene:SWCNTs attained at various current densities ranging from 0.0125 to 25 mA cm$^{-2}$. **Figure S6** reports the CD curves of WJM-graphene and WJM-graphene:SWCNTs obtained at various current densities. From CD curve data, the $C_{areal}$ of the MSCs can be calculated by the equation $C_{areal} = (|i|t_d)/(A_{geom}\Delta V)$,[159,167,168] where $|i|$ is the module of the applied current (i), $t_d$ is the discharge time, $A_{geom}$ is the geometrical area of the devices (calculated by including the area of interdigitated fingers and the interspaces between them) and $\Delta V$ is the voltage range of the measurement. As shown in Figure 5c, $C_{areal}$ values increased with the addition of SWCNTs to the WJM-graphene as active material. More in detail, $C_{areal}$ of 1.324 mF cm$^{-2}$ (5.296 mF cm$^{-2}$ for single electrode) was obtained for WJM-graphene:SWCNTs at a current density of 0.0125 mA cm$^{-2}$, which was 3.43-fold higher than that of WJM-graphene (0.385 mF cm$^{-2}$, 1.54 mF cm$^{-2}$ for single electrode). Lastly, by adopting PG as current collectors, high $C_{areal}$ (> 0.140 mF cm$^{-2}$) was observed even when operated at high CD rates, up to current density of 25 mA cm$^{-2}$. At 0.2 mA cm$^{-2}$, PG/WJM-graphene:SWCNTs have shown a 2.88-fold increase of $C_{areal}$ (equal to 1.033 mF cm$^{-2}$) compared to that obtained for WJM-graphene:SWCNTs (0.359 mF cm$^{-2}$). The inset of Figure 5c shows the $C_{vol}$ of the tested MSCs (calculated by $C_{vol} = (|i|t_d)/(Vol\Delta V)$, where Vol is the volume of the devices taking into account the whole thickness of the interdigitated fingers, including PG-based current collectors when used) at various current densities. At the lowest current density (0.0125 mA cm$^{-2}$), $C_{vol}$ was 0.112 F cm$^{-3}$ (0.448 F cm$^{-3}$ for single electrode) and 0.490 F cm$^{-3}$ (1.961 F cm$^{-3}$ for single electrode) for WJM-graphene and WJM-graphene:SWCNTs, respectively. These values approach the state-of-the art achieved by ultrathin MSCs using tens of nm-thick film of reduced graphene oxide as active materials ($C_{areal}$ up to 17.9 F cm$^{-3}$[75]).[32,75] At low current density (< 0.1 mA cm$^{-2}$), the use of PG as current collector decreased the $C_{vol}$ of PG/WJM-graphene:SWCNTs compared to that of WJM-graphene:SWCNTs. This is attributed to the overall thickness of the interdigitated electrodes, which also include the 10 μm-thick PG paper. However, the use of PG paper still allowed to achieve remarkable $C_{vol}$ (up to 0.362 F cm$^{-3}$ at current density of 0.0125 mA cm$^{-2}$), especially at high CD current densities (> 0.25 mA cm$^{-2}$).

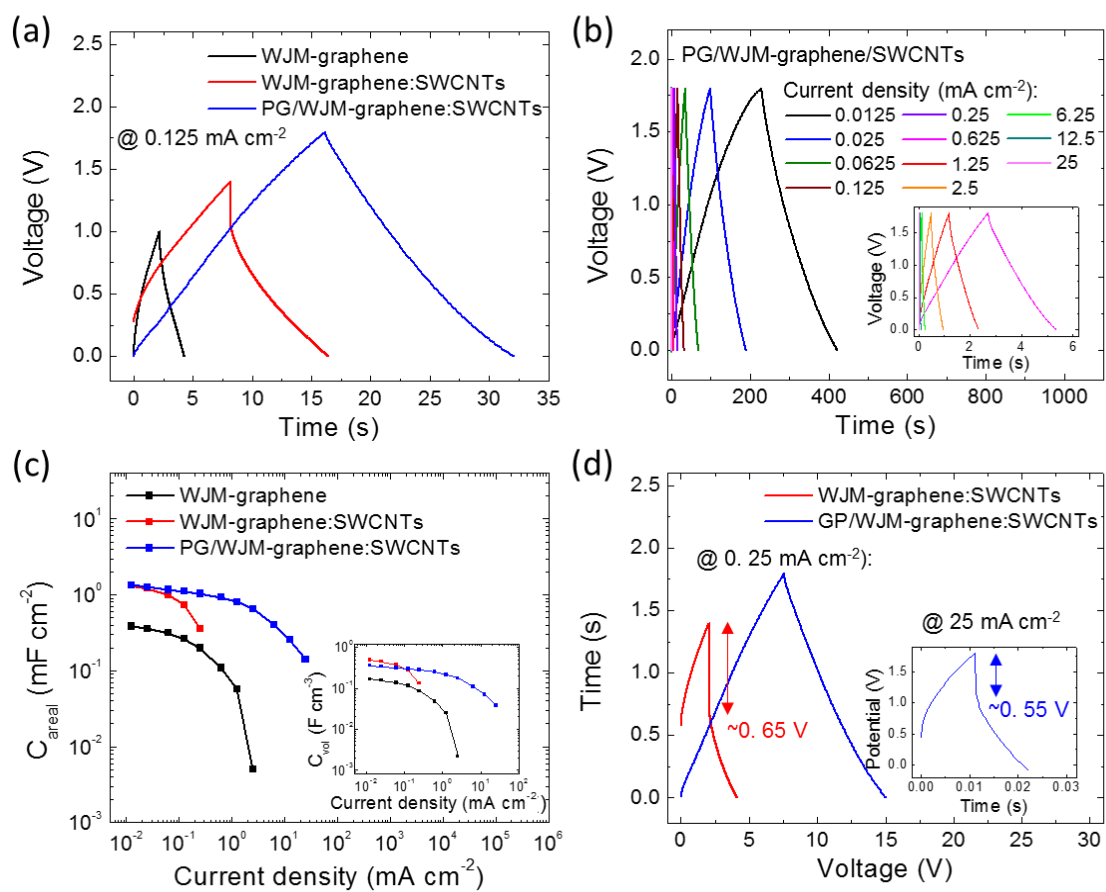

**Figure 5. Capacitance evaluation of screen-printed MSCs.** (a) Comparison between galvanostatic CD curves of WJM-graphene (black), WJM-graphene:SWCNTs (red) and PG/WJM-graphene:SWCNTs (blue) at a current density of 0.125 mA cm$^{-2}$. (b) Galvanostatic CD curves of PG/WJM-graphene:SWCNTs at different current density (from 0.0125 mA cm$^{-2}$ to 25 mA cm$^{-2}$). (c) Value of the $C_{areal}$ plotted as a function of the current density for WJM-graphene (black), WJM-graphene:SWCNTs (red) and PG/WJM-graphene:SWCNTs (blue). The inset shows the values of the $C_{vol}$ plotted as a function of the current density different for the same MSCs. (d) Comparison of the galvanostatic CD curves WJM-graphene:SWCNTs (red) and PG/WJM-graphene:SWCNTs (blue) at a current density of 0.25 mA cm$^{-2}$. The inset panel shows the galvanostatic CD curve of the PG/WJM-graphene:SWCNTs at current density of 25 mA cm$^{-2}$. The $V_{drop}$ values, which were used for ESR estimation, are also indicated for each CD curve.

In these experimental conditions, WJM-graphene:SWCNTs have shown a resistive behaviour. The capacitive losses observed with increasing the current density were ascribed to the potential drops of the equivalent series resistance (ESR) of the MSCs.[8,9,159,169,170] The ESR can be calculated by the galvanostatic CD curves, using the equation ESR = $V_{drop}/|i|$, where $V_{drop}$ is the voltage drop at half-cycle of the CD curve.[8,9,159,167-169] As shown in Figure 5d, in absence of PG paper as current collector, the in-plane electrical resistance of the interdigitated electrodes mainly contributed to the overall ESR, whose value (~3.2 kΩ) is two orders of magnitude higher than that of PG/WJM-graphene:SWCNTs (~22 Ω). Therefore, depending on the final requirement in terms of volume and operating power, the use of PG paper as current collector can be assessed whether advantageous. It is worth noting that the as-obtained volumetric performance allows to achieve outstanding $C_{areal}$ by adopting higher mass loading (milligram-scale)/thickness (ten micrometres-scale) of active materials compared to competing MSC technologies (ten micrograms-/ten nanometres-scale).[15-19,32,75] Consequently, PG/WJM-graphene:SWCNTs achieved a power density of 20.13 and 1.13 mW cm$^{-2}$ at energy density of 0.064 and 0.361 μWh cm$^{-2}$. **Figure 6** shows the Ragone plot obtained for PG/WJM-graphene:SWCNTs, together with some working point (power density, energy density), which are results of graphene-based MSCs reported in literature.[40,75,171,172,173] Although extraordinary volumetric performance have been previously reported for graphene-based MSCs,[40,75,172] the corresponding areal performance, in terms of both energy density and power density, were typically inferior to those achieved by our PG/WJM-graphene:SWCNTs. Notably, record high areal performance have been demonstrated by using carbon nano-onions-based MSCs,[40] in which, however, the synthesis of carbon nano-onions starts from expensive nanodiamond powder and require high-temperature process exceeding 1700 °C.[40]

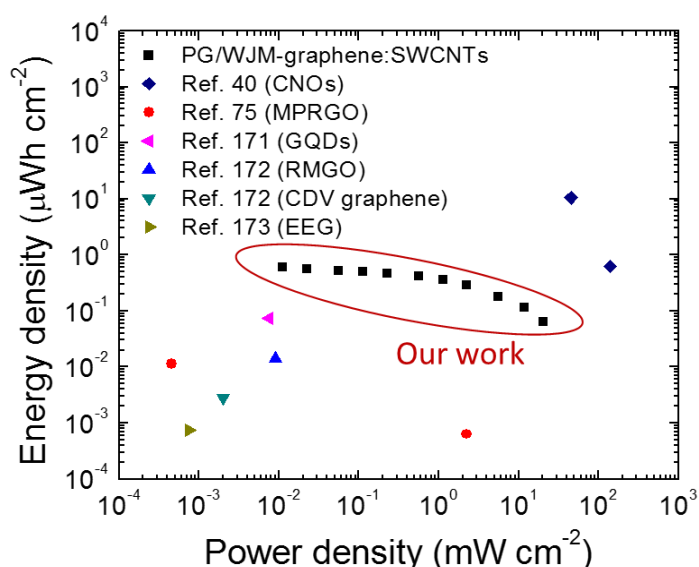

**Figure 6. Ragone plot of screen-printed PG/WJM-graphene:SWCNTs.** The comparison of working point (power density, energy density) with some relevant graphene-based MSCs reported in literature are also shown, demonstrating that PG/WJM-graphene:SWCNTs held superior areal performance. In the legend, the active material of MSCs used in literature is reported: CNOs (carbon nano-onions),[40] MPRGO (methanol-plasma reduced graphene oxide),[75] GQDs (graphene quantum dots),[171] RMGO (reduced multilayer graphene oxide),[172] CVD graphene (graphene synthetized via chemical vapour deposition)[172] and EEG (electrochemically exfoliated graphene).[173]

In order to prove the durability and the mechanical properties of the as-produced MSCs, galvanostatic CD cycling was carried out over 10000 cycles (**Figure 7**a) and different bending-type stresses with different radii of curvature (R) and bending angle (θ) (Figure 7b). As shown in Figure 7a, PG/WJM-graphene:SWCNTs shows excellent cycling stability. In fact, up to 98% of capacitance was retained after 10000 cycles at a current density of 0.1875 mA cm$^{-2}$. Notably, the high coulombic efficiency (stabilized above 96% at 0.1875 mA cm$^{-2}$, and approaching to 99% at current density higher than 0.5 mA cm$^{-2}$, Figure 5b) indicates that the double-layer formation was highly reversible, without the occurrence of parasitic faradaic reactions. Similar results were also obtained for WJM-graphene, as shown in the inset of Figure 7a. The high-durability of the devices can be attributed to the in-plane interdigitated graphene-based structure which allows for a favourable ultrafast uptake of the flow of electrolyte ions into or removal from the graphene layers in a short diffusion pathway.[75,76,172]

Figure 7c shows the capacitance retention plots over bending with R of 1 cm for PG/WJM-graphene:SWCNTs. The inset panel of Figure 7c shows the same test with R equal to 2 cm. For R = 1 cm, the device retained more than 97% of the initial capacitance, with coulombic efficiency exceeding 95%. For a R of 2 cm, a negligible capacitance loss (~1%) was observed with a coulombic efficiency > 95%. The devices were also tested at different θ ranging from 0° (unbent) to 90° and 180° (folded) (Figure 7d). After folding at both these θ values, the device exhibited a 6% increase of the capacitance, without significant variation of the coulombic efficiency. The increase of capacitance can be tentatively attributed to a favourable mesoscopic rearrangement of the active materials in the folded electrodes.[174]

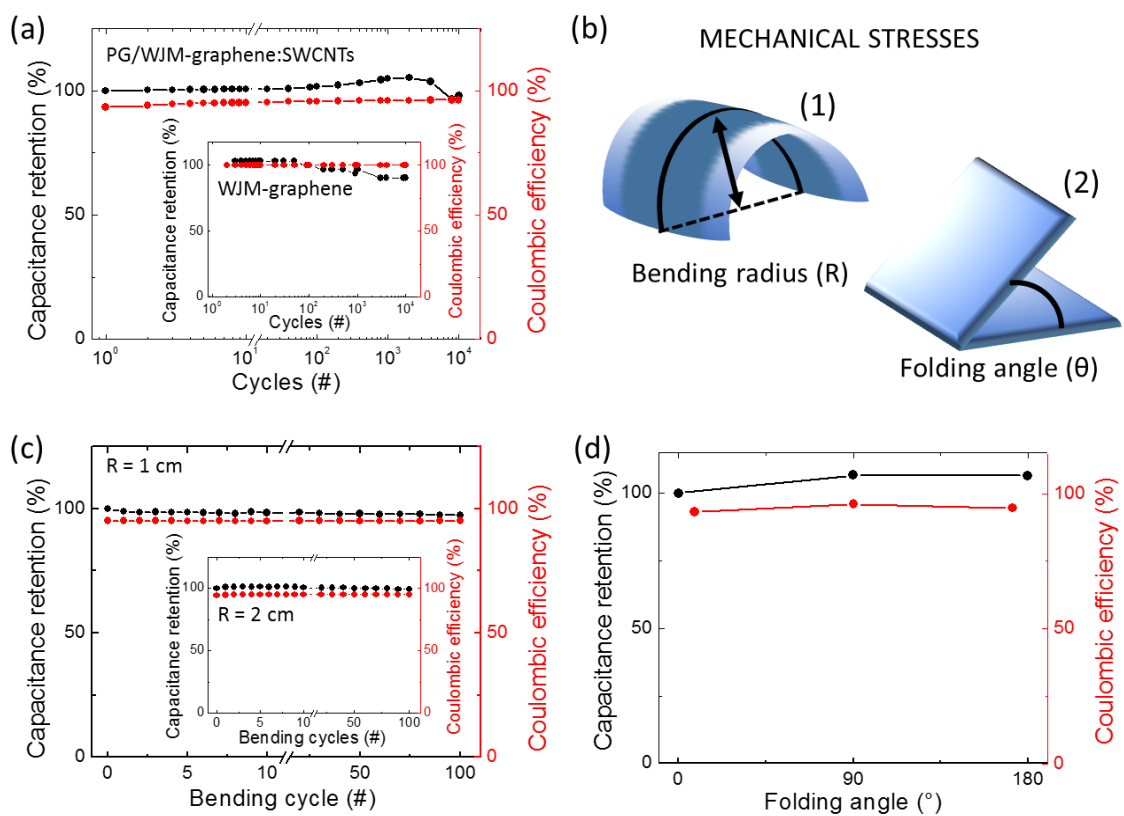

**Figure 7. Durability and mechanical flexibility of screen-printed MSCs.** (a) Capacitance retention of PG/WJM-graphene:SWCNTs over 10000 CD cycles at current density of 0.125 mA cm$^{-2}$. (b) Schematic illustration of the adopted mechanical stresses: (1) bending and (2) folding. (c) Capacitance retention of PG/WJM-graphene:SWCNTs over 100 bending at R of 1 and 2 cm (inset panel) (black, left y-axis). The coulombic efficiency is also plotted as a function of the bending cycles (red, right y-axis). (d) Capacitance retention of folded PG/WJM-graphene:SWCNTs at θ of 90° and 180° (black, left y-axis). The coulombic efficiency is also plotted as a function of the folding angle (red, right y-axis).

For wearable application, waterproof property is also crucial,[13,175,176] because wearable devices often face wet environment such as sweat and rain, as well as laundry cycles.[177,178] Therefore, washability test was also carried out on PG/WJM-graphene:SWCNTs. Before undergoing a washing cycle, the device was encapsulated with cross-linked ethylene vinyl acetate (EVA) *via* hot sealing process at 140 °C. Notably, EVA is an encapsulant benchmark with both waterproof properties and stress-crack resistance (*i.e.*, mechanical flexibility).[179,180] For the waterproof test, both CV and CD curves were measured after a washing cycle in home-laundry conditions. In order to simulate practical home-laundry conditions, the device was accommodated into a microfleece garment (**Figure 8**a) and the washing cycle was carried out by using both detergent and softener at 60 °C, followed by centrifugation at 1200 rpm to remove water excess and fasten the laundry drying. As shown in Figure 8b, similar current densities were recorded by CV measurements before and after the washing cycle. The occurrence of parasitic faradaic reaction in the voltage range of 0–1.8 V might be ascribed to the sealing process, which could partially alter the hydrogel-polymer electrolyte. The optimization of the sealing method is currently under investigation. As obtained from CD curves recorded at 0.1 mA cm$^{-2}$ (inset of Figure 8b), the device exhibited a C$_{areal}$ of 1.21 and 1.17 mF cm$^{-2}$ in the voltage windows of 0–1.4 V and 0–1.8 V, respectively. Notably, these values are comparable to those measured on reference device before sealing and washing processes (1.10 mF cm$^{-2}$ at current density of 0.1 mA cm$^{-2}$).

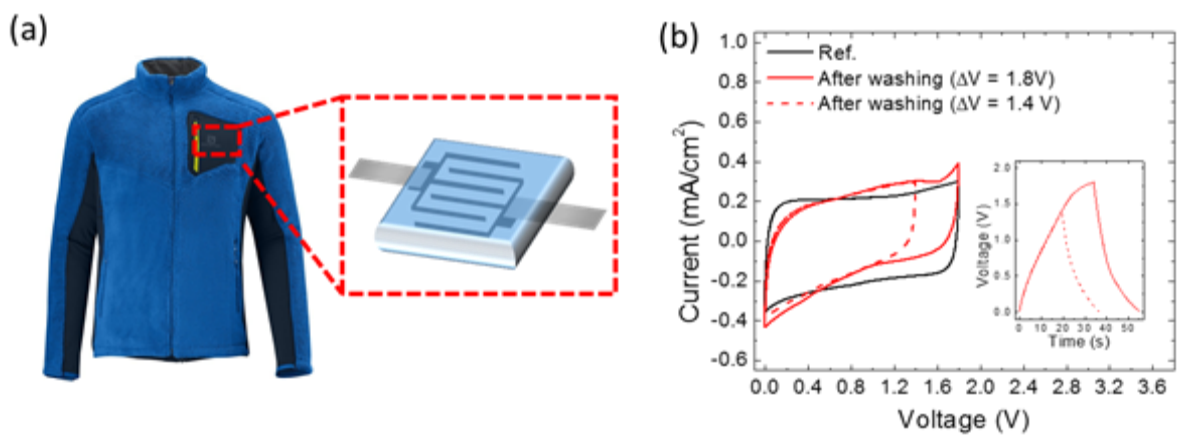

**Figure 8. Washability tests of EVA-sealed screen-printed MSCs.** (a) Schematic illustration of the MSCs accommodation into microfleece garment simulating practical home-laundry conditions. (b) Electrochemical characterization (CV and CD measurements) of the MSCs after washing cycle using both detergent and fabric softener at 60 °C, followed by centrifugation at 1200 rpm.

## 3. Conclusion

In this work, we exploited a scalable production of graphene inks *via* wet-jet milling (WJM) exfoliation and solvent-exchange processes for screen-printed, flexible, solid-state and washable micro-supercapacitors (MSCs). In particular, WJM exfoliation of graphite in N-methyl-2-pyrrolidone allowed the production of large volumes (litre-scale) of concentrated high quality (single-/few-layers graphene) graphene dispersions (gram per litre-scale) in short times (second-scale) to be achieved. The subsequent solvent-exchange process was effective for the formulation of a screen printable WJM-graphene-based ink in water/ethanol (70:30) and terpineol (1% wt.). Such ink was then used to fabricate flexible solid-state MSCs on plastic substrate. The addition of single-walled carbon nanotubes dispersed in ethanol to WJM-graphene ink and the use of pyrolytic graphite paper as current collector were used for increasing the active film porosity and decrease the equivalent series resistance of the MSCs, respectively. The optimized MSCs exhibited an areal capacitance ($C_{areal}$) up to 1.324 mF cm$^{-2}$ (5.296 mF cm$^{-2}$ for single electrode), which corresponds to a volumetric capacitance ($C_{vol}$) of 0.490 F cm$^{-3}$ (1.961 F cm$^{-3}$ for single electrode). The screen-printed MSCs also displayed high-rate performance (voltage scan rate > 10 V s$^{-1}$ and charge/discharge –CD– current density up to 25 mA cm$^{-2}$), which allowed the MSCs to operate up to power densities of 20.13 and 1.13 mW cm$^{-2}$ at energy density of 0.064 and 0.361 µWh cm$^{-2}$, respectively. The devices exhibited excellent electrochemical and mechanical performances over CD cycling (10000 times), bending cycles (100 times under bending radius of 1 cm), and folding (up to angles of 180°). Moreover, waterproof cross-linked ethylene vinyl acetate-encapsulated MSC retained its supercapacitor performance after a home-laundry cycle, thus providing washable properties for prospective wearable electronics. These results establish an effective route to scale-up the production throughput of graphene-based MSCs, compatibly with an industrial-like production of graphene as solution processable active material for supercapacitors.

## 4. Experimental Section

*Exfoliation of graphite in NMP and debundling of SWCNTs.* A mixture of 200 g of bulk graphite flakes (+100 mesh from Sigma Aldrich) and 20 L of NMP (Sigma Aldrich) was prepared. The mixture was placed in the container and mixed with a mechanical stirrer (Eurostar digital Ika-Werke). The graphite exfoliation was carried out through WJM exfoliation method, by adopting a three-pass protocol.[112] The first WJM pass used a nozzle aperture of 0.30 mm. The piston-pass, defined as the number of times the piston was charged and discharged with graphite/NMP mixture, was set to 1000 passes (10 mL per pass). The processed sample was then collected in a second container. The WJM process was then repeated passing the sample through the 0.15 mm nozzle. Finally, the nozzle was changed to 0.10 mm diameter and a third WJM step was carried out. The as-produced sample was named WJM-graphene. The SWCNTs were prepared by dispersing as-purchased SWCNTs (CheapTube) in EtOH at a concentration of 10 g L$^{-1}$ using ultrasonic bath-assisted method. Experimentally, the SWCNTs dispersion in EtOH was sonicated for 30 min using a horn probe (Vibra-cell 75185, Sonics), with vibration amplitude set to 45%. The sonic tip was pulsed at a rate of 5 s on and 2 s off. An ice bath was used during sonication in order to minimize heating effects.

*Formulation of screen printable pastes.* The as-produced WJM-graphene ink was dried using a rotary evaporator (Heidolph, Hei-Vap Value), at temperature of 70 °C and pressure of 5 mbar. Then, the as-obtained powder was re-dispersed in a mixture of $H_2O$/EtOH (70:30) and terpineol (1 wt%) (solvent-exchange process) with a concentration of 75 g $L^{-1}$. The addition of SWCNTs (25 wt%) into WJM-graphene ink was also evaluated as active material ink with a total material concentration of 80 g $L^{-1}$.

*Preparation of hydrogel-polymer electrolyte.* 1 g of $H_3PO_4$ was added into 10 mL of deionized water, and then 1 g of PVA (molecular weight: 89 000–98 000, Sigma-Aldrich) was added. The whole mixture was heated to 80 °C under stirring, obtaining $H_3PO_4$-doped PVA-based hydrogel-polymer electrolyte.

*Micro-supercapacitor fabrication.* A screen printer was used to fabricate the interdigitated electrodes. The screen printing setup consisted in a 58 × 48 $cm^2$ 55T (threads per cm) aluminium frame, using polyester as a mesh material. The 55T aluminium frame had a single mesh opening of 0.1 mm, yielding a maximum resolution of 0.3 mm. The printed configuration consisted of six fingers (1 mm thick) forming the interdigitated structure (interspace between fingers of 600 μm), providing an active area of 1 $cm^2$. The printing strategy involved multiple printing passes until proper mass loading of active material was reached (1.5 mg for WJM-graphene and 1 mg for WJM-graphene:SWCNTs). The substrate was 1 μm-thick PET foil (Sigma Aldrich). Depending on the printed paste, the MSCs named WJM-graphene and WJM-graphene:SWCNTs were obtained. Electrodes of 10 μm-thick PG paper (PGS, Panasonic) with the same interdigitated shape of screen-printed electrodes were also evaluated as current collectors for WJM-graphene:SWCNTs-based MSCs. The corresponding MSCs were named PG/WJM-graphene:SWCNTs. After printing the active materials, the structures were dried at 80 °C under vacuum for 1 h to remove solvents. Finally, $H_3PO_4$-doped PVA-based electrolyte was drop-casted onto the printed structure and dried in ambient conditions overnight. Moreover, washability test was also carried out on PG/WJM-

graphene:SWCNTs. For this purpose, the device was properly encapsulated with crosslinked EVA, *via* hot sealing process at 140 °C.

*Characterization of materials.* Transmission electron microscopy images were taken with a JEM 1011 (JEOL) transmission electron microscope, operating at 100 kV. Morphological and statistical analyses were carried out by using ImageJ software (NIH) and OriginPro 9.1 software (OriginLab), respectively. The samples were prepared by drop casting WJM-graphene dispersions onto ultrathin C-film onto holey carbon 400 mesh Cu grids (Ted Pella Inc.). The WJM-graphene dispersion was diluted 1:50 before their deposition. The grids were stored under vacuum at room temperature to remove the solvent residuals.

The AFM measurements were carried out by using a NanoWizard III AFM system (JPK Instruments, Berlin) in intermittent contact mode. PPP-NCHR cantilevers (Nanosensors, USA) with a nominal tip diameter of 10 nm and a drive frequency of ~320 kHz were used. AFM images (512×512 data points) of 5×5 $\mu m^2$ were collected by keeping the working set point above 70% of the free oscillation amplitude. The scan rate for acquisition of images was 0.7 Hz. Height profiles were processed by using the JPK Data Processing software (JPK Instruments, Germany) and the data were analysed with OriginPro 9.1 software. Statistical analysis was carried out by means of Origin 9.1 software on multiple AFM images for each sample. The samples were prepared by drop-casting 1:30 diluted WJM-graphene dispersion in NMP onto freshly-cleaved mica sheets (G250-1, Agar Scientific Ltd., Essex, U.K.) and dried under vacuum.

Raman spectroscopy measurements were carried out by using a Renishaw microRaman invia 1000 using a 50× objective (numerical aperture of 0.75), with an excitation wavelength of 514.5 nm and an incident power on the samples of 5 mW. The samples were prepared by drop casting the 1:30 diluted WJM-graphene dispersion in NMP onto a Si wafer covered with 300 nm thermally grown $SiO_2$ (LDB Technologies Ltd.). The bulk graphite was analysed in the powder form. For each sample, 50 spectra were collected. OriginPro 2016 was used to perform the deconvolution and statistics.

X-ray photoelectron spectroscopy analysis was accomplished using a Kratos Axis Ultra$^{DLD}$ spectrometer. The sample was obtained by drop casting WJM-graphene dispersion in NMP onto Au-coated silicon wafers. The samples were then dried at 200 °C overnight to remove solvent residuals. The XPS spectra were acquired using a monochromatic Al K$_\alpha$ source operating at 20 mA and 15 kV. The analysis was carried out on a 300 μm × 700 μm area. High-resolution spectra of C 1s and Au 4f peaks were collected at pass energy of 10 eV and energy step of 0.1 eV. Energy calibration was performed setting the Au 4f$_{7/2}$ peak at 84.0 eV. Data analysis was carried out with CasaXPS software (version 2.3.17).

*Characterization of MSCs.* Scanning electron microscopy analysis was performed using a Helios Nanolab® 600 DualBeam microscope (FEI Company) and 5kV and 0.2 nA as measurement conditions. The SEM images of the interdigitated electrodes were collected without any metal coating or pre-treatment.

Specific surface area measurements of MSC electrodes were carried out in Autosorb-iQ (Quantachrome) by Kr physisorption at temperatures of 77 K. The SSA was calculated using the multipoint BET model,[181] considering equally spaced points in the P/P$_0$ range of 0.009 – 0.075. P$_0$ is the vapour pressure of Kr at 77 K, corresponding to 2.63 Torr.

The electrochemical measurements were performed with potentiostat/galvanostat (VMP3, Biologic). Cyclic Voltammetry measurements were carried out at various voltage scan rate, ranging from 0.01 to 20 V s$^{-1}$. Galvanostatic CD measurements were at different current density, ranging from 0.0125 to 25 mA cm$^{-2}$. All the electrochemical measurements were carried out at room temperature. For the waterproof test, CV and CD curves were recorded after a washing cycle in home-laundry machine (IWC 71251, Indesit). In order to simulate practical home-laundry conditions, the device was accommodated into microfleece garment (High Pile Fz Midlayer M Bl/Bl, Salomon) and the washing cycle was carried out using both detergent (Liquid Multicolor, Dixan®, Henkel) and fabric softener (Cura 3D, Perlana®, Henkel) at 60 °C, followed by centrifugation at 1200 rpm to remove water excess and fasten the laundry drying.

**Acknowledgements**
Sebastiano Bellani and Elisa Petroni contributed equally to this work. This project has received funding from the European Union's Horizon 2020 research and innovation program under grant agreement 785219–GrapheneCore2. We thank Electron Microscopy facility – Istituto Italiano di Tecnologia for support in TEM data acquisition; Nanochemistry facility, Istituto Italiano di Tecnologia for support in material characterization.

# Supporting Information

**Scalable production of graphene inks *via* wet-jet milling exfoliation for screen-printed micro-supercapacitors**

*Sebastiano Bellani, Elisa Petroni, Antonio Esau Del Rio Castillo, Nicola Curreli, Beatriz Martín-García, Reinier Oropesa-Nuñez, Mirko Prato and Francesco Bonaccorso\**

**Description of wet-jet milling apparatus**

As schematically illustrated in Figure 1a of the main text, the wet-jet milling (WJM) process exploits high pressure (180–250 MPa) generated by a hydraulic piston to force the passage of the solvent/layered-crystal mixture through perforated disks, with adjustable hole diameters (0.3–0.1 mm, named nozzle), located into the processor. More in detail, the processor consists in a set of five different perforated and interconnected disks, which form a set of channels (**Figure S1**).

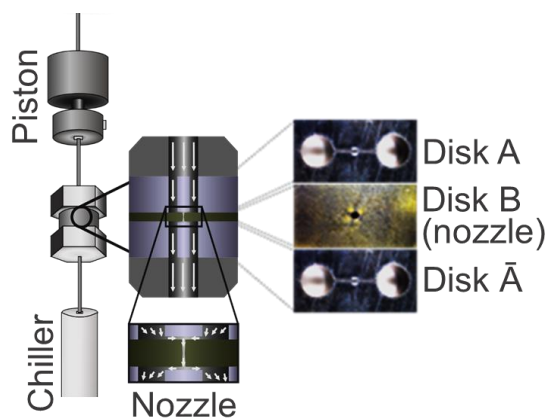

**Figure S1.** Close-up view of the processor in the wet-jet milling apparatus. The zoomed images show the channels configuration and the disks arrangement. The white arrows indicate the fluid path. On the right side, a top view of the holes and channels on each disk. The disks A and Ā have two holes of 1 mm in diameter, separated by a distance of 2.3 mm from centre to centre and joined by a half-cylinder channel of 0.3 mm in diameter. The thickness of the A and Ā disks is 4 mm. The disk B consists of a 0.10 mm nozzle and is the core of the system. The thickness of the B disk is 0.95 mm.

The configuration of the disks divides the flow in two streams (Disk A). Subsequently, solvent and layered material mixture passes through the nozzle (a perforated hole with sub-mm diameter, Disk B). Here, shear forces are formed and promote the sample exfoliation.[1,2]

**Raman spectroscopy analysis**

A typical Raman spectrum of graphene shows, as fingerprints, G, D and 2D peaks.[3,4] The G peak, positioned at ~1585 cm$^{-1}$, corresponds to the E$_{2g}$ phonon at the Brillouin zone centre.[5] The D peak is due to the breathing modes of sp$^2$ rings and requires a defect for its activation by double resonance.[1,2,6,7,8] Double resonance happens as intra-valley process, *i.e.,* connecting two points belonging to the same cone around K or K'.[5124,9] This process gives rise to the D' peak for defective graphene.[5,10,11] The 2D peak is the second order of the D peak.[5] The 2D peak is always seen, even in the absence of D peak, since no defects are required for the activation of two phonons with the same momentum, one backscattered from the other.[10] Moreover, 2D peak is a single peak centred at ~2680 cm$^{-1}$ for single-layer graphene,[5] whereas is a superposition of multiple components, the main being the 2D$_1$ and 2D$_2$ components, for few-layers graphene.[3-5] In graphite, the intensity of the 2D$_2$ band is twice the 2D$_1$ band,[3,4,12] For single-layer graphene the 2D band is roughly four times more intense than the G peak.[3,4] Multi-layers graphene (> 5 layers) exhibits a 2D peak, which is almost identical, in term of intensity and lineshape, to the graphite case.[3,4] Few-layers graphene (< 5 layers), instead, has a 2D$_1$ peak more intense than the 2D$_2$.[13] Taking into account the intensity ratios of the 2D$_1$ and 2D$_2$, it is possible to estimate the flake thickness.[4,13]

**Figure S2** reports the Raman spectroscopy analysis of the measurements performed on the WJM-produced sample (WJM-graphene). Figure S2a shows that the ratio between the intensity of D and G peak –I(D)/I(G)– ranges from 0.1 to 1.2, whereas approaches to null values in graphite (see Section 2.1). The position of G –Pos(G)– and the full width half maximum of G –FWHM(G)– range from 1578 to 1583 cm$^{-1}$ (Figure S2b) and from 14 to 25 cm$^{-1}$ (Figure S2c), respectively. The plot of I(D)/I(G) *vs*. FWHM(G) (Figure S2d) shows the absence of any correlation, which means that the WJM process did not induce in-plane defects. The plot of the intensity of 2D$_1$ –I(2D$_1$)– *vs*. the intensity of 2D$_2$ –I(2D$_2$) (normalized on I(G) gives information on the flakes thickness (see Figure S2e). The dashed line I(2D$_1$)/I(G) = I(2D$_2$)/I(G) in Figure S2e roughly represents the multilayer condition (~5 layers). Thus, graphitic samples whose I(2D$_2$)/I(2D$_1$) value falls above the aforementioned line can be assumed to be less than 5 layers thick, while for the ones that the aforementioned value falls below, are considered thicker than 5 layers, thus displaying graphite-like properties.

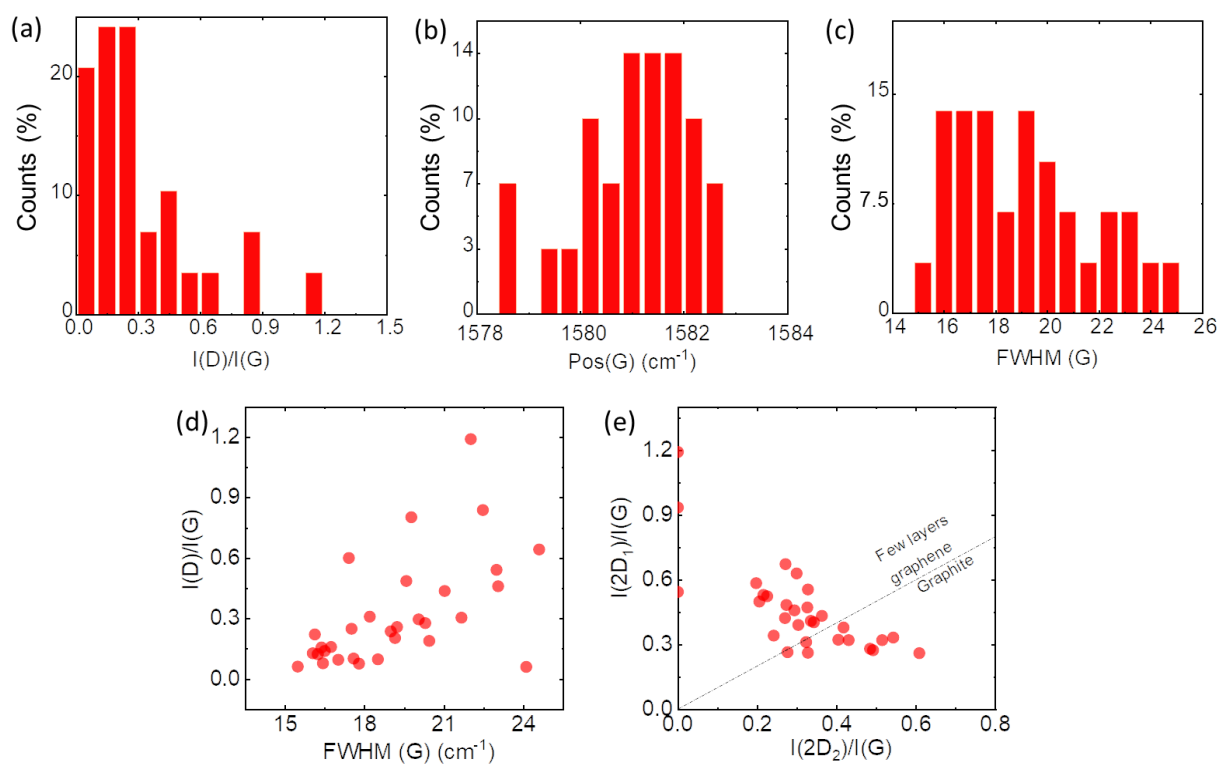

**Figure S2. Raman spectroscopy measurements analysis for WJM-graphene.** Statistical analysis of: (a) I(D)/I(G). (b) Pos(G) and (c) FWHM(G). (d) Plot of I(D)/I(G) *vs*. FWHM(G). (e) Plot of I(2D$_1$)/I(G) *vs*. I(2D$_2$)/I(G). The dashed line I(2D$_1$)/I(G) = I(2D$_2$)/I(G), representing the multilayer condition (~5 layers), is also shown.

**Scanning electrode microscopy of WJM-graphene film**

**Figure S3** shows a top-view SEM image of a WJM-graphene film in absence of SWCNTs.

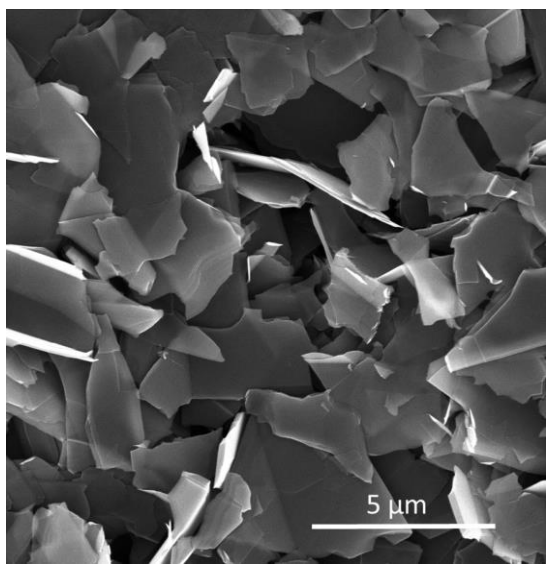

**Figure S3.** Top-view SEM image of a representative WJM-graphene film.

**Evaluation of the voltage windows for the micro-supercapacitor**s

The working voltage range of the produced micro-supercapacitors (MSCs) was evaluated by cyclic voltammetry (CV) measurements at low scan rate (*i.e.*, 10 mV s$^{-1}$). In this experimental condition, the capacitive current density, which is proportional to the voltage scan rate, is reduced, and parasitic faradaic reaction contribution can be discriminated. **Figure S4** shows the CV curves measured at voltage scan rate of 10 mV s$^{-1}$ for the various MSCs (see notation in the main text), *i.e.*, WJM-graphene (Figure S4a), WJM-graphene:SWCNTs (Figure S4b) and PG/WJM-graphene:SWCNTs (Figure S4c).

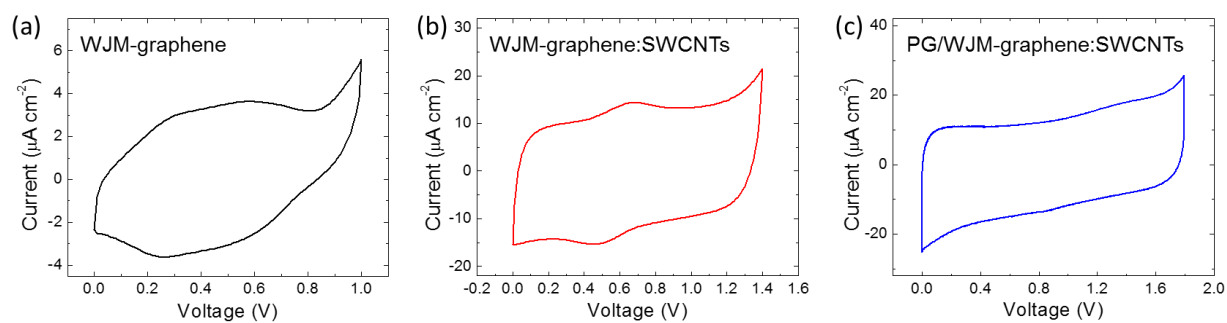

**Figure S4.** Cyclic voltammetry curve of (a) WJM-graphene, (b) WJM-graphene:SWCNTs and (c) PG/WJM-graphene:SWCNTs at a voltage scan rate of 10 mV s$^{-1}$.

Since nearly-rectangular CV shapes and the absence of redox peaks indicated that the MSCs show double-layer capacitive behaviour, the upper voltages were limited to 1, 1.4 and 1.8 V for WJM-graphene, WJM-graphene:SWCNTs and PG/WJM-graphene:SWCNTs, respectively. Above these voltage values, the parasitic faradaic reactions started, as evidenced by whisker-like CV shapes.

**Supplementary cyclic voltammetry measurements**

**Figure S5** reports the CV measurements for WJM-graphene (Figure S5a,b) and WJM-graphene:SWCNTs (Figure S5c,d) at voltage scan rate ranging from 0.01 to 20 V s$^{-1}$. Differently from PG/WJM-graphene:SWCNTs, the devices exhibited an increasing resistive behaviour following the increase of the voltage scan rate.

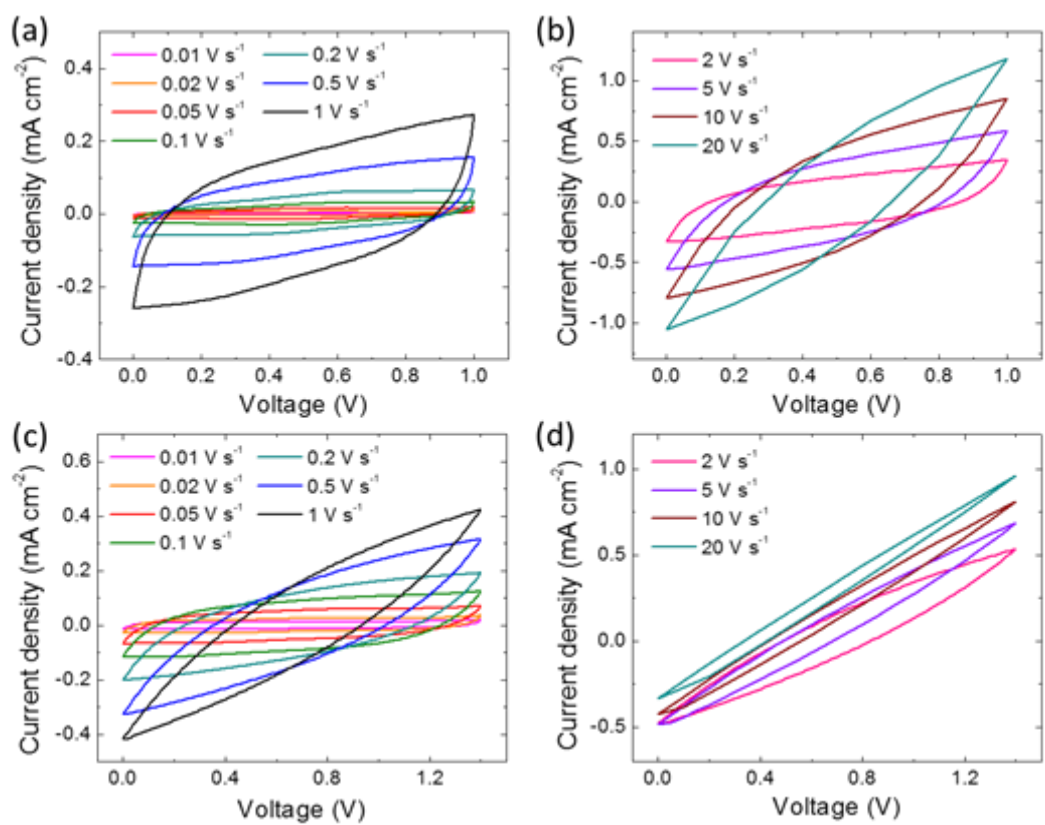

**Figure S5. Cyclic voltammetry characterization of WJM-graphene and WJM-graphene:SWCNTs.** (a,b) Cyclic voltammetry curves of WJM-graphene at different voltage scan rate (from 0.01 V s$^{-1}$ to 1 V s$^{-1}$ for panel a, from 2 V s$^{-1}$ to 20 V s$^{-1}$ for panel b). (c,d) Cyclic voltammetry curves of WJM-graphene:SWCNTs at different voltage scan rate (from 0.01 V s$^{-1}$ to 1 V s$^{-1}$ for panel c, from 2 V s$^{-1}$ to 20 V s$^{-1}$ for panel d).

As explained in the main text (see Section 2.2, Figure 4 and Figure 5d), this effect was ascribed to the equivalent series resistance (ESR) of the interdigitated electrode made of solely active material without PG paper as current collector.

**Supplementary galvanostatic charge/discharge measurements**

**Figure S6** shows the galvanostatic charge/discharge (CD) curve of WJM-graphene (Figure S6a) and WJM-graphene:SWCNTs (Figure S5b) obtained at various current densities (from 0.0125 to 25 mA cm$^{-2}$). From CD curves, the $C_{areal}$ of the MSCs, (see Figure 5c, Section 2.2) were calculated.

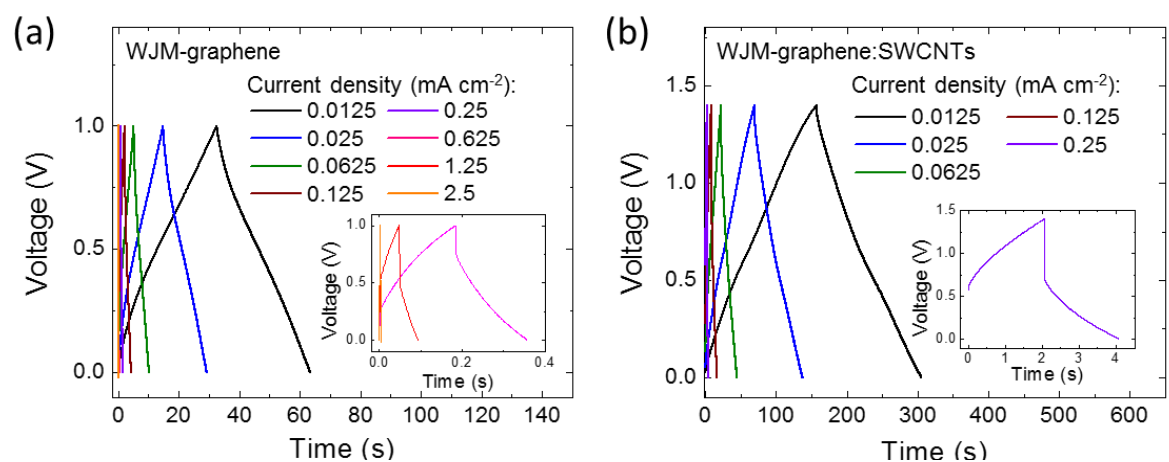

**Figure S6. Galvanostatic CD measurements for WJM-graphene and WJM-graphene:SWCNTs.** (a) Galvanostatic CD curves of WJM-graphene at various current density (from 0.0125 mA cm$^{-2}$ to 2.5 mA cm$^{-2}$). The inset shows the temporal enlargement of the CD curve obtained at current density of 0.625, 1.25 and 2.5 mA cm$^{-2}$. (b) Galvanostatic CD curves of WJM-graphene:SWCNTs at various current densities (from 0.0125 mA cm$^{-2}$ to 2.5 mA cm$^{-2}$). The inset shows the temporal enlargement of the CD curves obtained at current density of 0.625, 1.25 and 2.5 mA cm$^{-2}$.